\documentclass[12pt,a4paper]{article}
\usepackage[a4paper,margin=1in]{geometry}
\usepackage{times}
\usepackage{setspace}
\usepackage{hyperref}
\usepackage{titlesec}
\usepackage{graphicx}
\usepackage{amsmath, amssymb}
\usepackage{booktabs} 

\titleformat{\section}{\large\bfseries}{\thesection.}{0.5em}{}
\titleformat{\subsection}{\normalsize\bfseries}{\thesubsection.}{0.5em}{}

\title{Einstein Was Not a Flat Physicalist: Principle Theories, Constructive Theories, and the Direction of Constraint}

\medskip

\author{Galina Weinstein\thanks{The Department of Philosophy, University of Haifa.}}
\date{}

\begin{document}
\maketitle
\doublespacing

\begin{abstract}
Einstein’s distinction between \emph{principle theories} and \emph{constructive theories} is methodological rather than metaphysical. \emph{Principle theories} such as thermodynamics and relativity articulate empirically distilled constraints that delimit admissible microphysical models, while \emph{constructive theories} remain provisional and revisable. This paper reconstructs Einstein’s framework from primary sources and argues that recent appeals to it by Meir Hemmo and Orly Shenker—under the banner of \emph{Flat Physicalism}—invert its functional hierarchy. What is presented as an Einsteinian template instead supports a reductionist metaphysics foreign to Einstein’s methodology and increasingly misaligned with the structural commitments of contemporary physics.
\end{abstract}

\section{Introduction}

Einstein’s distinction between \emph{principle theories} and \emph{constructive theories}, articulated most explicitly in his 1919 essays and later reflections, has long been understood as a methodological device governing modes of theory construction rather than as a metaphysical hierarchy. Thermodynamics and relativity exemplify this approach: they articulate empirically grounded constraints—impossibilities and invariances—that delimit the space of admissible physical models, while leaving open a plurality of constructive realizations. The asymmetry Einstein insists upon is methodological and epistemic: \emph{principle theories constrain; constructive theories explore within those constraints}.

This interpretation has been carefully reconstructed in the historical and philosophical literature by Jeroen van Dongen \cite{VD}, Marco Giovanelli \cite{G-1, G-2}, Don Howard \cite{How}, Gerald Holton \cite{Holton, Hol73}, Michel Janssen \cite{Jan}, Martin J. Klein \cite{Klein}, and others. Nonetheless, a persistent tension arises in contemporary philosophy of physics. A body of work by Orly Shenker,%
\footnote{According to Shenker’s CV (updated December 2020), she is a Member of the Academic Committee of the Hebrew University’s Einstein Archive.} 
Meir Hemmo and collaborators extensively appeal to Einstein’s terminology while assigning it a fundamentally different role. Across discussions of \emph{Flat Physicalism}, statistical mechanics, the Past Hypothesis, the arrow of time, and mental causation, Einstein’s distinction is redeployed as a metaphysical ordering: \emph{principle theories} are treated as high-level phenomenological summaries, while \emph{constructive theories} are accorded ontological primacy. The vocabulary remains Einsteinian, but the direction of constraint is reversed.

Clarifying this tension requires separating three claims that are often run together under the heading of the principle–constructive distinction. The first is \emph{methodological}: principle theories articulate empirically refined constraints on admissible models. The second is \emph{epistemic}: within their proper domains, such theories enjoy exceptional stability because they are grounded in broad empirical regularities rather than speculative micro-ontologies. The third is \emph{metaphysical}: that constructive theories represent a deeper or more fundamental level of reality, while principle theories are merely surface-level descriptions.

Einstein endorses the first claim and, with explicit qualifications, the second. He does not endorse the third. His distinction is a division of epistemic labor, not a hierarchy of ontological levels. Principle theories are not explananda awaiting microphysical reduction; they function as restrictive frameworks that discipline and guide constructive theorizing.

By contrast, the contemporary ``Einsteinian template'' employed by Shenker, Hemmo, and their co-authors tacitly relies on the third claim while presenting itself as grounded in the first two. Principle theories are treated as explananda rather than as constraints, and constructive theories are assigned ontological primacy. The central thesis of this paper is that this template depends on a systematic role reversal: it preserves Einstein’s terminology while assigning to his distinction a function his texts explicitly deny.

The aim of this paper is twofold. First, I reconstruct Einstein’s principle–constructive distinction from the primary sources, emphasizing its constraint-based logic and epistemic asymmetry. Second, I analyze how its contemporary deployment in defenses of \emph{Flat Physicalism} inverts this structure and supports a classical reductionist program that is neither licensed by Einstein’s methodology nor well aligned with the conceptual commitments of modern physics. More broadly, the paper addresses the standards governing appeals to historical authority in contemporary metaphysics and philosophy of physics.

\section{Einstein’s Principle–Constructive Distinction}
\emph{The 1919 essay: Relativity as a principle theory}

In late 1919, following the British eclipse expeditions that confirmed the light-bending prediction of general relativity, Albert Einstein agreed to write an explanatory article for \emph{The Times} of London. Written in German as ``Was ist Relativitätstheorie?'' and published in English as ``Time, Space, and Gravitation'' \cite{Ein19-1, Ein19-2}, the article was intended not merely as popularization, but as a methodological clarification of the kind of theory relativity is.

In the essay, Einstein contrasts \emph{constructive theories} (\emph{konstruktive Theorien}) with \emph{principle theories} (\emph{Prinziptheorien}) \cite{Ein19-1}. This distinction is not merely classificatory but methodological and epistemological in character \cite{Holton}. According to Holton, Einstein regarded this distinction as central to understanding both the creative logic of his own work and the limitations of unconstrained model-building in physics.

\emph{Constructive theories} aim to build complex phenomena from hypothetical microscopic constituents. Their explanatory strategy is bottom-up and model-driven: from assumed elements and laws, they seek to reconstruct observed behavior. Einstein’s canonical example is the kinetic theory of gases, which explains macroscopic thermal phenomena by positing molecular motion. Such theories are often intuitively powerful and \emph{anschaulich} (vivid or pictorial), but their microscopic assumptions remain conjectural and permanently revisable \cite{Ein19-1, Ein19-2}.

As Holton stresses, Einstein did not reject constructive theorizing as such. He himself produced notable \emph{constructive theories}, most famously the light quantum hypothesis, which guided the explanation of the photoelectric effect published in 1905. Yet Einstein consistently regarded such constructions as heuristic devices: empirically effective, closely tied to specific phenomena, but provisional and conceptually secondary. In Holton’s reading, Einstein viewed early quantum theory precisely in this light—not as a new fundamental ontology or a principle-level framework, but as an indispensable stopgap, lacking the axiomatic distance from experience required for a foundational theory \cite{Holton, Hol73}.

\emph{Principle theories} proceed in the opposite direction. Rather than positing underlying mechanisms, they begin with empirically established, highly general constraints—principles that express what is physically impossible or invariant—and derive consequences analytically. Thermodynamics is Einstein’s paradigmatic case. From the impossibility of perpetual motion, it deduces universal restrictions on all physical processes, without reliance on any particular molecular hypothesis. Its strength lies not in mechanistic detail but in the security, generality, and unifying power of its principles \cite{Ein19-1}.

Einstein explicitly classifies relativity as a \emph{principle theory}. Like thermodynamics, it begins with empirically grounded constraints—the relativity principle and the constancy of the speed of light—which delimit the permissible form of all physical laws. Relativity does not hypothesize a microstructure of matter or an ether; it imposes invariance requirements that any admissible \emph{constructive theory} must satisfy. In this sense, relativity plays the same methodological role as thermodynamics: it restricts the space of possible physical theories rather than constructing one from speculative elements \cite{Ein19-1}.

``Special relativity thus imposes a kinematical constraint on all dynamical laws,'' as Janssen emphasizes, ``So, \emph{even as a constructive theory}, special relativity plays the heuristic role of providing constraints on further theorizing. This heuristic role was the important feature of principle theories for Einstein. He did not introduce the principle–constructive distinction to put labels on already established theories. He was essentially characterizing different strategies for finding new ones—develop a concrete model (the constructive strategy) or find constraints on such models (the principle strategy)'' \cite{Jan}.
Understood in this way, Einstein’s classification of relativity as a \emph{principle theory} does not deny the legitimacy of constructive work; it fixes the direction of methodological constraint. \emph{Constructive theories} explore possibilities, but \emph{principle theories} delimit them.

Holton emphasizes that Einstein regarded this constraint-based, non-constructive role of relativity as decisive. Only in \emph{principle theories}, Einstein believed, are the postulated axioms sufficiently removed from immediate experience and sufficiently insulated from \emph{ad hoc} adjustment to support a genuinely creative conceptual framework—one capable of organizing the totality of facts rather than merely accommodating isolated phenomena \cite{Holton}. Their power lies not in modeling mechanisms but in imposing structural constraints.

Einstein further stresses that \emph{principle theories} possess a distinctive logical structure. They are logically tight: from a small number of fundamental principles, universal consequences follow. If any derived consequence fails empirically, the theory as a whole must be abandoned. This vulnerability is inseparable from their epistemic authority. \emph{Constructive theories}, by contrast, permit piecemeal modification of hypotheses without global collapse \cite{Ein19-1}.

\emph{Thermodynamics, restrictive principles, and the avoidance of a chaos of possibilities}

Einstein’s retrospective reflections in the ``Autobiographical Notes'' sharpen this methodological contrast \cite{Ein49}. There, thermodynamics occupies a privileged epistemic position. Its foundation is formulated not as a descriptive generalization but as a modal prohibition: ``the laws of nature are such that it is impossible to construct a perpetuum mobile (of the first or second kind).'' Einstein characterizes this as an \emph{einschränkendes Prinzip}, a restrictive principle that excludes vast classes of conceivable physical laws.

The autobiographical context is revealing. Einstein describes his early encounter with mathematics as a Buridan’s-ass situation: confronted with a proliferation of specialized domains and lacking reliable criteria for identifying what is fundamental, he found himself paralyzed by choice \cite{Ein49}. The methodological lesson he later drew in physics—namely, that progress requires sharply formulated principles capable of restricting the space of admissible theories—can be read as a mature response to this epistemic predicament.

Einstein is explicit about the physics. Reflecting on his early work, he writes that he gradually ``despaired of the possibility of discovering the true laws by constructive efforts based on known facts ... Only the discovery of a general formal principle could lead us to secure results'' \cite{Ein49}. As Holton notes, this despair was not directed at empirical inquiry as such, but at unconstrained constructive modeling as a route to fundamental understanding \cite{Holton, Hol73}. What Einstein sought were principles sufficiently general, formal, and independent of particular mechanisms to yield \emph{gesicherte Ergebnisse}—secured results.

Special relativity is explicitly aligned with thermodynamics. Lorentz invariance functions as an \emph{einschränkendes Prinzip} for the laws of nature, directly comparable to the prohibition of perpetual motion. Many different \emph{constructive theories} may satisfy the same principle; the principle itself fixes the admissible form of all of them \cite{Ein49}. Such principles are not deduced from micro-models but are \emph{abgelauscht}—``eavesdropped on nature''—by identifying invariant structural features that persist across disparate phenomena \cite{Ein14}. In Holton’s formulation, they therefore constitute the most stable and least revisable stratum of physical knowledge \cite{Holton, Hol73}.

This methodological asymmetry helps illuminate episodes in Einstein’s own practice that might otherwise appear to challenge it. During the development of general relativity—most notably in the \emph{Entwurf} phase—Einstein temporarily lacked a sufficiently strong principle constraint and consequently wandered among competing constructions. Yet this dual strategy was \emph{instrumental in deepening Einstein’s understanding}: constructive exploration proceeded, but genuine progress resumed only once stronger principle constraints, above all general covariance, crystallized. Constructive work was not abandoned; it was disciplined by principle \cite{VD, G-1}.
Klein’s analysis of Einstein’s engagement with thermodynamics reinforces this point from a different angle. Klein emphasizes that thermodynamic principles function for Einstein as uniquely secure constraints that guide and delimit constructive theorizing across domains, rather than as explananda awaiting reduction to microphysics \cite{Klein}. The role of thermodynamics in Einstein’s thought is precisely to prevent theoretical inquiry from dissolving into what Einstein himself called a \emph{Chaos der Möglichkeiten} \cite{CPAE8}, Doc. 111, a ``chaos of possibilities.'' Episodes of intensive constructive activity, whether in gravitation or statistical physics, that exemplify the conditions under which principle-level constraints become indispensable \cite{G-1}.

\emph{Methodological significance}

Across the 1919 essay \cite{Ein19-1, Ein19-2}, the ``Autobiographical Notes'' \cite{Ein49}, and other methodological reflections—including the 1933 Herbert Spencer lecture \cite{Ein33}—Einstein’s position \emph{is consistent} once the distinction between practice and priority is kept in view. Einstein repeatedly affirmed the indispensability of constructive work, the creative freedom involved in concept formation, and the role of mathematical invention. \emph{What he did not do was to reverse the direction of constraint}. Even where constructive efforts dominate in practice, principles remain irreducible, indispensable, and not themselves explananda.

The hierarchy at issue is therefore methodological rather than metaphysical. \emph{Constructive theories} propose hypothetical microstructures and remain permanently revisable. \emph{Principle theories} articulate empirically grounded constraints that delimit what any acceptable \emph{constructive theory} may assert. Principles do not replace models; they guide and discipline their construction. Any reconstruction that treats \emph{principle theories} as dispensable phenomenology and \emph{constructive theories} as the sole bearers of ontology \emph{abandons the structure Einstein explicitly articulated}.

\section{Flat Physicalism and the Einsteinian Distinction}

Orly Shenker, Meir Hemmo, and their collaborators repeatedly appeal to Einstein’s writings— most notably ``Time, Space, and Gravitation'' and the \emph{Autobiographical Notes} \cite{Ein19-2, Ein49}—as methodological touchstones for their work on \emph{Flat Physicalism}, statistical mechanics, and the philosophy of mind. While these citations are generally accurate at the level of quotation, systematic difficulties arise from the way Einstein’s \emph{principle/constructive} distinction is put to work. What functions in Einstein as a methodological asymmetry governing theory construction is redeployed as a two-level metaphysical schema in which \emph{principle theories} are downgraded to high-level phenomenology and \emph{constructive theories} are assigned ontological primacy. The vocabulary remains Einsteinian, but the direction of constraint is reversed.

The present critique does not oppose reductive microphysical research as such. The issue is whether Einstein’s distinction can legitimately be invoked to underwrite a ``flat,'' micro-ontological metaphysics without explicit revision. Across a wide range of contexts— thermodynamics, the arrow of time, physicalism, mental causation, and quantum foundations—the same structural pattern recurs: Einstein’s terminology is retained, but the functional role of the distinction is inverted.

Against the Einsteinian background reconstructed in the previous section, the point of contrast can be stated succinctly. In Einstein’s framework, \emph{principle theories} articulate empirically refined constraints that delimit the space of admissible models, while \emph{constructive theories} explore candidate mechanisms within those bounds. In the work of Hemmo and Shenker, by contrast, \emph{principle theories} are treated as provisional summaries of regularities to be explained away by deeper microphysical accounts, while \emph{constructive theories} are treated as the sole bearers of ontology. This inversion is not local or accidental; it is supported by a series of arguments that appear distinct but are structurally unified.

The treatment of the \emph{Past Hypothesis} provides a clear illustration. Standard Boltzmannian statistical mechanics, combined with time-reversal invariant microdynamics and typicality reasoning conditioned on the present macrostate, yields the \emph{minimum entropy theorem}: a typical microstate is overwhelmingly likely to exhibit entropy increase both toward the future and toward the past. This result conflicts with thermodynamics, memory, and empirical evidence, generating a \emph{skeptical catastrophe} in David Albert’s sense \cite{A}. To block this conclusion, statistical mechanics is supplemented by the \emph{Past Hypothesis}, postulating a low-entropy boundary condition in the past.

Dan Baras and Shenker analyze whether this \emph{Past State} calls for explanation \cite{BS}. They do not themselves propose a physical explanation, but what matters here is the explanatory framework they presuppose. Thermodynamics is treated as the \emph{explanandum}, statistical mechanics as the constructive \emph{explanans}, and the \emph{Past State} as the remaining unexplained element. Statistical mechanics is described as a ``high-level'' \emph{constructive theory}, with thermodynamics as a corresponding \emph{principle theory}.

This way of speaking already departs from Einstein’s framework. For Einstein, \emph{constructive theories} are not ``high-level''; they are bottom-up, micro-structural, and \emph{anschaulich}. Nor does Einstein treat thermodynamics as surface phenomenology awaiting explanation. His distinction concerns modes of theorizing, not ontological depth. The reductionist hierarchy presupposed in the \emph{Past Hypothesis} literature reflects later reconstructions of statistical mechanics rather than Einstein’s methodological outlook.

A parallel inversion appears in Hemmo and Shenker’s account of the \emph{psychological arrow} of time \cite{HS202}. Rejecting the \emph{Albert–Loewer Mentaculus program}, they argue that appealing to a global low-entropy boundary condition does not genuinely explain temporal experience, but merely postulates it as a fundamental constraint \cite{HS202, HS22-1}. Since entropy gradients are not intrinsically time-directed, and there is no credible account of how brains could register them as such, the Past Hypothesis cannot ground the psychological arrow. Instead, they propose that the \emph{psychological arrow} arises from a local, effective symmetry-breaking within the brain itself, with other arrows—including the thermodynamic arrow—defined only relative to this psychologically fixed direction.

This move has two consequences. First, it renders the Second Law observer-dependent, in direct tension with Einstein’s treatment of thermodynamics as an observer-independent \emph{principle theory}. Second, it sits uneasily with \emph{Flat Physicalism} itself. A macro-level asymmetry in certain subsystems is permitted to orient global thermodynamic and cosmological claims, thereby introducing a top-down dependence that is incompatible with a strictly ``flat'' ontology.

The same structure reappears in Shenker’s treatment of the foundations of statistical mechanics \cite{S-171}. Although Einstein’s terminology is retained, statistical mechanics is presented as the attempt to apply mechanics, together with probability measures, to explain thermodynamic phenomena. Thermodynamic quantities are systematically identified with mechanical aspects or functions of microstates. Constraint flows from micro to macro, rather than from principle to constructive. Shenker appeals to Howard’s analysis \cite{S-172}; however, Howard’s account treats the principle–constructive distinction as a methodological tool \cite{How}, not a metaphysical hierarchy of ontological levels. 

This inversion underpins broader metaphysical claims. In reformulating \emph{Hempel’s dilemma}, Erez Firt, Hemmo, and Shenker identify instability in any ontology tied to a revisable \emph{deep structure} \cite{FHS}. But this dilemma arises only because ontology is located exclusively at the constructive level \cite{HS23}: ``deep structure theory – that explains the 'principle theory'.'' In Einstein’s framework, revisability at the constructive level does not undermine the determinacy or authority of principle-level constraints. The instability is generated by the inversion itself.

The same template drives their critique of \emph{Davidson’s anomalous monism}. Davidson’s position affirms \emph{token identity} without \emph{type identity}: every \emph{mental event} is physical, but mental predicates do not form law-governed physical kinds \cite{David}. Hemmo and Shenker reconstruct \emph{anomalous monism} as a \emph{principle theory} standing above a constructive base of events and descriptions, modeled on statistical mechanics \cite{HS-201, S-15}. Once so reconstructed, \emph{anomalous monism} is forced into a dilemma between \emph{reductive physicalism} and \emph{token-dualism} \cite{HS-22}. This dilemma does not arise from Davidson’s commitments; it is produced by assigning anomalous monism a role it was never meant to play.

\emph{Flat physicalism} is introduced as the unifying resolution of these instabilities \cite{HS22}. Ontological layering is rejected altogether. Only the microphysical level is real; all higher-level kinds are fixed by the complete microstate of the world. Thermodynamics, psychology, and special-science regularities are reconstructed as patterns in coarse-grainings of microstates. The generalized \emph{Hempelian dilemma} is avoided by ``flattening'' ontology completely \cite{HS22, S-173}.

The coherence of this position, however, depends on abandoning Einstein’s constraint-based asymmetry. Thermodynamics and relativity are reduced to descriptive summaries; constructive microphysics is assigned full ontological authority. Einstein’s vocabulary is retained, but its function is reversed.

This reversal becomes especially clear in the treatment of \emph{Maxwell’s Demon}. Hemmo and Shenker argue that Demons are compatible with classical mechanics and statistical mechanics, provided entropy is defined over observer-relative macrostates \cite{HS-12}. Different phase-space partitions yield different thermodynamic regularities. The Second Law becomes a contingent pattern rather than a principled impossibility. Yet Einstein’s claim that thermodynamics \emph{will never be overthrown} is explicitly restricted to the domain in which its concepts apply. Either Demon scenarios fall outside that domain, or Einstein’s principle-theoretic conception is rejected. What is not tenable is the suggestion that Einstein’s framework already licenses the Demon-friendly conclusion.

A further instance of the same role reversal appears in Amit Hagar and Hemmo’s critique of Jeffrey Bub’s information-theoretic interpretation of quantum mechanics \cite{HH}. Bub explicitly models his approach on Einstein’s \emph{principle theories}: quantum mechanics is treated as a theory of structural constraints on possibilities. At the same time, \emph{constructive interpretations} are subordinate attempts to fit dynamics within that framework \cite{B}. Hagar and Hemmo argue that such a \emph{principle theory} is explanatorily incomplete and must be supplemented by constructive dynamics. In doing so, they again elevate \emph{constructive theories} to epistemic primacy, treating the principle level as structurally insufficient rather than constraint-giving.

Finally, developments in quantum statistical mechanics undermine the classical reduction template on which \emph{Flat Physicalism} depends. \emph{Entanglement, decoherence, scrambling, and universality} reveal macroscopic kinds as structurally defined equivalence classes relative to algebras of observables, not as functions of microstates \cite{Lashkar, Zurek}. Shenker herself acknowledges these features, yet attempts to preserve ``flatness'' by privileging the global quantum state  \cite{S-21}. Once translated into genuinely quantum terms, this strategy collapses: macrostates are not functions of microstates but structurally robust patterns with explanatory autonomy.

None of this threatens \emph{physicalism}. It undermines only a classical \emph{Flat Physicalism}. A structural physicalism—one that recognizes principle-level constraints and higher-level invariants as physically real—is both compatible with modern physics and faithful to Einstein’s methodological insights.

The recurring difficulty, therefore, is not error but equivocation. Einstein’s authority is repeatedly invoked while the functional role of his distinction is reversed. One may defend a ``flat,'' reductionist metaphysics, but one cannot plausibly present it as continuous with Einstein’s conception of \emph{principle theories}. Einstein’s distinction is a division of epistemic labor, not a template for downgrading principles into dispensable summaries. The vocabulary survives; the structure does not.

\section{Conclusion}

The aim of this paper has not been to contest physicalism, nor to deny the legitimacy of reductive microphysical research. It has been to clarify the methodological logic of Einstein’s principle–constructive distinction and to assess the extent to which contemporary appeals to that distinction remain faithful to its original function.

Once Einstein’s framework is reconstructed with sufficient care, a consistent asymmetry emerges. \emph{Principle theories} articulate empirically grounded constraints that delimit the space of admissible laws, while \emph{constructive theories} explore candidate mechanisms within those bounds. This asymmetry concerns modes of theorizing, not levels of being; it governs the direction of constraint in theory construction rather than assigning ontological primacy.

Against this background, the appeal to Einstein in defenses of \emph{Flat Physicalism} becomes problematic. Across a wide range of contexts—statistical mechanics, the Past Hypothesis, the psychological arrow of time, anomalous monism, and the foundations of quantum theory—the same structural pattern recurs. Einstein’s terminology is retained, but the functional role of the distinction is inverted: principle theories are treated as provisional summaries to be explained away by deeper microphysical accounts, while constructive theories are assigned exclusive ontological authority. Constraint flows upward from microstructure rather than downward from principles.

This inversion is not a local interpretive choice but a systematic commitment. It underwrites the rejection of genuine autonomy in the special sciences, motivates the flattening of ontology into a single microphysical level, and drives the reconstruction of thermodynamic and psychological asymmetries as observer-relative artifacts of coarse-graining. Yet this strategy succeeds only by abandoning the very feature that made Einstein’s distinction philosophically productive—its capacity to impose non-derivative constraints on theory space.

The tension becomes especially clear once contemporary physics is taken seriously. In quantum statistical mechanics, macroscopic kinds are not functions of microstates but structurally defined equivalence classes relative to algebras of observables. Entanglement, decoherence, and universality reveal higher-level regularities that are robust under microphysical variation and explanatorily indispensable. These are not merely epistemic conveniences; they function as constraint structures that no purely “flat” microstate ontology can recover.

None of this threatens physicalism. What it undermines is a classical flat physicalism that collapses principles into summaries and treats microstates as the sole bearers of physical reality. A physicalism that is both responsive to modern physics and faithful to Einstein’s methodological insights must recognize the reality of principle-level constraints and the non-trivial structure they impose on admissible dynamics.

Einstein’s distinction was never intended as a license for ontological flattening. It was a division of epistemic labor that preserved the creative and unifying role of principles precisely by insulating them from \emph{ad hoc} constructive revision. One may defend a flat, reductionist metaphysics—but one cannot plausibly present it as continuous with Einstein’s conception of \emph{principle theories}. The vocabulary may survive the translation; the structure does not.

\end{document}